\documentstyle{l-aa}
\begin{document}
\thesaurus{03 (11.11.1; 11.19.6; 11.19.5; 11.01.1; 11.14.1; 11.09.1 NGC 2841)}
\title{Decoupled nuclei and nuclear polar rings in regular spiral
galaxies. NGC 2841
\thanks{Based on observations collected with the 6m telescope
of the Special Astrophysical Observatory (SAO) of the
Russian Academy of Sciences (RAS) which is operated under the
financial support of Science Department of Russia (registration number
01-43).}}
\author{O. K. Sil'chenko \inst{1}
\and V. V. Vlasyuk \inst{2} \and A. N. Burenkov \inst{2}}
\offprints{O. K. Sil'chenko}
\institute{
Sternberg Astronomical Institute, University av. 13,
Moscow 119899, Russia
\and
Special Astrophysical Observatory, Nizhnij Arkhyz, Karachaj-Cherkess
Republic, 357147 Russia}
\maketitle
% \markboth
% \maintitlerunninghead{Decoupled Nucleus in NGC 2841}
% \authorrunninghead{O. K. Sil'chenko et al.}
\begin{abstract}

The Sb galaxy NGC~2841 was observed at the 6 m telescope of SAO RAS
with the Multi-Pupil Field Spectrograph and at the 1 m telescope of SAO
RAS with the long-slit spectrograph.
An unresolved nucleus of NGC~2841 is shown to be chemically decoupled
both in magnesium and in iron with abundance break estimates of
0.36 dex for Mg and 0.6 dex for Fe; an abundance gradient in the bulge
is seen only in the magnesium index and is typical for early-type disk
galaxies. The rotation axis of the nuclear ionized gas in NGC~2841 is
perpendicular
to that of the central stellar population; an existence of a
bulge stellar component with decoupled rotation momentum in the radius
range 5\arcsec-12\arcsec\ is suspected. A possible scenario for the
origin of the unusual central structure in NGC~2841 is proposed.


\end{abstract}

\section{Introduction}

Chemically and kinematically decoupled galactic nuclei have been
known for some years, but mostly in elliptical galaxies
(Jedrzejewski \& Schechter 1988; Bender 1988; Bender \& Surma 1992;
Davies et al. 1993; Carollo \& Danziger 1994). Early hypothesis
related this phenomenon to a merger of a smaller stellar system.
Later when it became clear that decoupled nuclei in elliptical galaxies
may be nuclear disks (Surma \& Bender 1995), a variant of dissipative
merging was proved to be better.

Several years ago we claimed the existence of chemically decoupled
nuclei in spiral galaxies for the first time (Sil'chenko et al. 1992).
We had obtained equivalent width radial profiles up to several
arcseconds from the center by measuring the MgI$\lambda5175$ absorption
line in three early-type spiral galaxies, NGC~615, 7013, and 7331, where
earlier we had detected dynamically decoupled nuclei, namely, strong
central mass concentrations (Afanasiev et al. 1989). It seemed
that in the central parts of their bulges any magnesium-strength
gradients were absent inside the error limits; but the nuclei
were decoupled, showing a higher magnesium-line strength at the level
of $2\sigma$ and more. The values of metallicity differences between
the decoupled nuclei and the surrounding bulges ranged from a factor
of two to almost an order of magnitude. Obviously a dissipationless
merging was
to be excluded in the case of decoupled nuclei in spiral galaxies.

We have begun a systematic search for chemically distinct nuclei in
galaxies by composing a list of galaxies with photometrically distinct
nuclei. Among 234 early-type galaxies with detailed multi-aperture
photoelectric data which were found in the catalogues of Longo and
de Vaucouleurs (1983, 1985) about 25\% of ellipticals and lenticulars
and about 50\% of early-type spirals show nuclei which are
prominently redder that the neighbouring bulges (Sil'chenko 1994).
So it may be possible that chemically distinct nuclei are a rather
common phenomenon. We have selected 34 objects of the northern sky
for a detailed investigation. As dust concentration may also provide
the red colour of the nuclei, bidimensional spectroscopy
was necessary to prove the dynamical and chemical distinctness of
nuclei in these galaxies. So we started an observational program
with the Multi-Pupil Field Spectrograph (MPFS) of the 6 m telescope
(Special Astrophysical Observatory, Nizhnij Arkhyz, Russia). Now
we have completed and published results on two galaxies from this
list. The elliptical galaxy NGC~1052 appears to possess a resolved
chemically decoupled core with a radius of 3\arcsec\ (300 pc);
the ionized gas inside this radius co-rotates with the stars
whereas the gas further outward rotates (and is distributed)
orthogonally to the stars
(Sil'chenko 1995). The small spiral galaxy NGC~4826 has become famous
during last three years by its counterrotating outer gaseous disk:
the sense of gas rotation suddenly changes at the radius of 1 kpc
(Braun et al. 1994; Rubin 1994; Rix et al. 1995). We have found
an unresolved chemically distinct nucleus in this galaxy; the central
gas rotates circularly and together with the stars (Sil'chenko 1996).
Though only two arbitrary examples may not be representative, some
suspicions have arisen that the phenomenon of chemically decoupled
nuclei can be related to external gas capture.

NGC~2841 was chosen as having a photometrically distinct red nucleus
(Sil'chenko 1994). But it has one
more property which is interesting for us. When we found the first three
chemically decoupled nuclei in spiral galaxies (Sil'chenko et al. 1992),
in all three early-type spiral galaxies possessing such nuclei the
orientation of the innermost isophotes was shown to be the same as that
of the outermost isophotes; in other words, their bulges look
axisymmetrical (Sil'chenko \& Vlasyuk 1992). And in NGC~2841 any turn
of the isophote major axis was thought to be absent too
(Mizuno \& Hamajima 1987)
so this galaxy was one of the best our candidates for possessing
a chemically distinct nucleus. Besides, it was known to have
a compact photometric nucleus whose brightness exceeded an extrapolation
of the de Vaucouleurs' bulge (Kormendy 1985).

\section{Observations and data reduction}

Spectral observations of the center of NGC~2841 were carried out in the
Special Astrophysical Observatory of RAS (Nizhnij Arkhyz, Russia) with
the Multi-Pupil Field Spectrograph (MPFS), installed in the prime
focus of the 6-meter telescope (Afanasiev et al. 1990), and also
by using a long-slit spectrograph, attached to the Cassegrain focus
of the 1-meter reflector of the Special Astrophysical Observatory.
The log of observations is presented in Table~1.

MPFS, which is a second (after CFHT TIGER system, see Bacon et al. 1995)
realisation of G. Courtes' (1982) concept of spatial sampling
of extended sources by means of a microlens array, has been
in active operation at the 6-meter telescope since 1989. Instruments of
this type are providing sufficient gain in investigations of nebulae and
galaxies with respect to a classical slit spectroscopy due to complete
coverage of studied sky area, independence of spectral
resolution on spatial resolution, absence of slit losses and of
the overall problem of object matching.

A set of enlargers, which project the object onto the lens array,
provide spatial sampling according to
seeing value --- from $\approx$0.3\arcsec\ to $\approx$1.7\arcsec\
per lens. Sizes of the used array varied between 9$\times$11
and 8$\times$16 elements.
We have chosen for our observations the 10$x$ enlarger,
providing a scale of 1.3--1.4\arcsec\ per lense and a field of view
of $\approx$10$\times$16\arcsec\ for different
instrument setups. The accurate scale values and sizes of field
of view are given in Table~1.

We have exposed two sets of spectra in the green spectral range
($\lambda\lambda$ 4700--5500\,\AA\AA)
with a spectral resolution near 6 \AA\ and one set of spectra in the red
wavelength range ($\lambda\lambda$ 6450--6700\,\AA\AA)
with a resolution about 2\,\AA.

To account accurately for the night sky background in the green spectral
range we have separately exposed the blank sky region
at 1.5\arcmin - 2\arcmin\ from the galaxy
with an exposure time of one half of that for the galaxy; the sky was
then (after spectrum extraction and linearization) smoothed
and subtracted.

The hollow-cathode lamps filled by He-Ne-Ar or Ne-Xe-Ar gas mixtures were
exposed before and after each exposure in order to provide
wavelength calibration of spectral data. Integrations of the twilight sky
were carried out for correcting system vignetting and
variations of the transmission by the individual lenses.

The long-slit spectrograph was installed at the Cassegrain focus
of the 1 meter Zeiss reflector and equipped by a fast (f/1.5)
Schmidt-Cassegrain camera and large-format
CCD (1040$\times$1160 elements). The four cross-sections at different
position angles were accumulated with spatial sampling
1.54\arcsec\ per pixel at a spectral resolution of 4\AA\
and over the spectral range of 4100-6000 \AA.
In order to clean spectral accumulations from cosmic ray hits
all exposures were divided in two.
For wavelength calibration we have used Ne-Xe-Ar lamps exposed
before and after each object exposure.

\begin{table*}
\caption[ ] {Spectral observations of NGC~2841}
% \begin{center}
\begin{flushleft}
\begin{tabular}{lcllllcl}
\hline\noalign{\smallskip}
Date & Telescope & Configuration & Exposure & Field & Scale
& PA of long side & Seeing \\
\hline\noalign{\smallskip}
20/21.10.93 & 6m & MPFS+IPCS $512\times512$ & 21 min &
$12\arcsec\times14\arcsec$ & 1.27\arcsec/lens & $152\degr$ &
1.5\arcsec \\
25/26.10.94 & 6m & MPFS+CCD $520\times580$ & 90 min &
$14\arcsec\times22\arcsec$ & 1.35\arcsec/lens & $-12\degr$ &
2\arcsec \\
12/13.01.96 & 1m & LS+CCD $1040\times1160$ & 60 min &
$4\arcsec\times3\arcmin$ & 1.54\arcsec/pixel & $152\degr$ &
3.7\arcsec \\
13/14.01.96 & 1m & LS+CCD $1040\times1160$ & 60 min &
$4\arcsec\times3\arcmin$ & 1.54\arcsec/pixel & $242\degr$ &
3\arcsec  \\
14/15.01.96 & 1m & LS+CCD $1040\times1160$ & 60 min &
$4\arcsec\times3\arcmin$ & 1.54\arcsec/pixel & $92\degr$ &
2\arcsec  \\
14/15.01.96 & 1m & LS+CCD $1040\times1160$ & 60 min &
$4\arcsec\times3\arcmin$ & 1.54\arcsec/pixel & $122\degr$ &
2\arcsec \\
27/28.02.96 & 6m & MPFS+CCD $1040\times1160$ & 60 min &
$17\arcsec\times22\arcsec$ & 1.43\arcsec/lens & $166\degr$ &
3.5\arcsec \\
\hline
\end{tabular}
% \end{center}
\end{flushleft}
\end{table*}

The basic data reduction steps -- bias subtraction, flatfielding,
cosmic ray hits removing, extraction
of one-dimensional spectra, transformation into wavelength scale,
construction of surface brightness maps and velocity fields --
were performed using the software developed
in the Special Astrophysical Observatory (Vlasyuk 1993).

The green spectra obtained with MPFS were used to construct
maps of surface brightness in the continuum at $\lambda 5100$
and two-dimensional velocity fields for the stellar component (only the
spectra obtained with CCD in 1996 are suitable for cross-correlation;
as a template, we have taken the spectrum of the nucleus).
Then the spectra of individual elements were added in the rings
centered on the nucleus;
the width of the rings was equal to the lens size
and the step between neighbouring rings was also equal to the lens
size.
These summarized spectra were used to derive radial dependencies
of absorption line strengths -- main attention was paid to H$\beta$,
MgI$b$, and Fe lines $\lambda 5270$ and $\lambda 5335$.

The red spectra were only used to derive a velocity field and
a spatial distribution of
the ionized gas: the long exposure has allowed us to measure weak
emission lines H$\alpha$ and [NII]$\lambda 6583$ in the center
of NGC~2841 for the first time.

Long-slit spectra obtained at the 1 m telescope were
tentatively destined to confirm the properties of the velocity fields
obtained with MPFS: as the slit was very broad, $s=4\arcsec$,
the measured picture
appears to be slightly smoothed, and there are problems with exact
localization of measured features.

From these data  we estimate an accuracy of stellar and gas velocities
as 20~$\mbox{km} \ \mbox{s}^{-1}$ and an accuracy
of absorption-line equivalent widths as 0.3\,\AA\ for the spectra
registered with IPCS and as 0.1\,\AA\ for the spectra registered with
CCD.

The benefit of the Multi-Pupil Field Spectrograph over the long-slit
spectrographs in the study of absorption line radial dependencies is
that after summing spectra in the rings, we deal with a radially
constant level of accumulated counts and do not need to apply
logarithmic binning to obtain the constant
equivalent-width accuracy over all the radii under consideration.

\section{Chemically decoupled nucleus in NGC~2841}

\begin{figure}
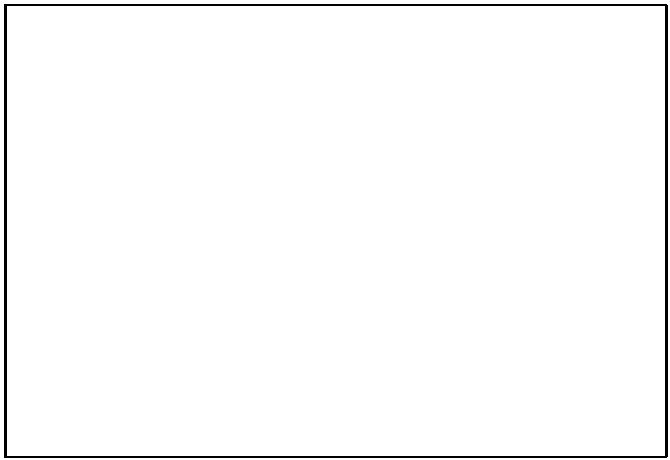

\picplace{6.0cm}
\caption[]{The magnesium-line radial dependence in the center of
NGC~2841.}
\end{figure}

Figure~1 presents radial profiles of the well-known absorption-line
index $Mgb$ (for the absorption-line index definitions --
see Worthey et al. 1994), measured in the center of NGC~2841 twice:
in 1993 and in 1996. As was mentioned above, the
indices measured with IPCS are far less accurate than those
measured with CCD, but in this case the two independent data sets
agree perfectly. The only discrepancy at $r=1.3\arcsec$ can be explained
by the seeing difference (see Table~1). Two conclusions can be derived
by examining Fig.~1: firstly, it is the first case during our
search for chemically decoupled nuclei where
a rather strong magnesium-strength
gradient in the bulge is present, and secondly, the unresolved nucleus
is decoupled by its increased Mg-line strength even if we extrapolate
the metallicity trend in the bulge towards $r=0$. The straight line
in Fig.~1 shows the least-square fit to the bulge magnesium-line trend in
the radius range 4\arcsec--14\arcsec. It is just the bulge, because
according to recent photometric data of Varela et al. (1996) the bulge
dominates in NGC~2841 up to $r\approx 50\arcsec$ (its $r_e=25\arcsec$).
The formula of the straight line in Fig.~1 is \\

\noindent
$Mgb = (3.66\pm0.25)-(0.088\pm0.031)r\arcsec$, \\

\noindent
so the nucleus which has $Mgb=4.47\pm0.06$ is distinguished from the
underlying bulge at a confidence level of more than $3\sigma$,
and the metallicity break
calibrated by using the models of old stellar populations from
Worthey (1994) is equal to 0.36 dex. By applying the same
metallicity-$Mgb$ relation to the $Mgb$ gradient in the bulge,
we obtain
$d\mbox{[m/H]}/d\mbox{log}r=-0.9\pm0.3$. Balcells \& Peletier (1994)
investigated
metallicity gradients in bulges of early-type disk galaxies by
considering profiles of the broad-band color $B-R$, and they
have reported a range of metallicity gradients from 0 to --1. So
our result for NGC~2841 is consistent with the mean characteristics
of early-type disk galaxies.

Here we must note that a linear $Mgb$ dependence on $r$ in the central
bulge is not physically conditioned -- it is only the most exact and
convenient presentation in this range of radius. For example, if we
apply a logarithmic relation, the fit would be less accurate by a
factor of two. As the structure of the bulge of NGC~2841 may be very
complex and inhomogeneous (see the next Section), we prefer not
to restrict ourselves by any physical model but rather to use
a good empirical relation.

\begin{figure}
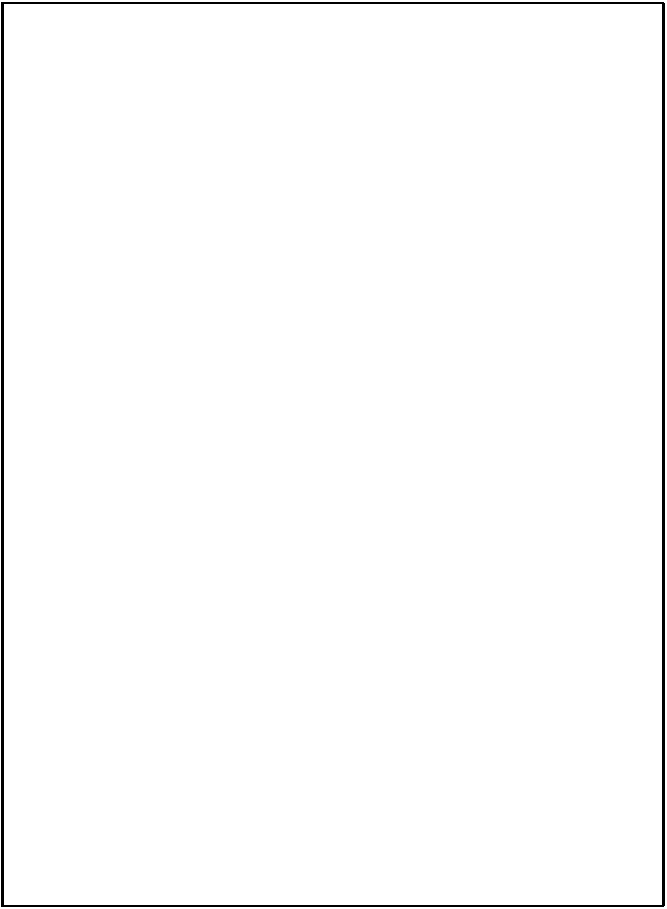

\picplace{12.0cm}
\caption[]{The Balmer absorption and iron-line radial dependencies
           in the center of NGC~2841.}
\end{figure}

Figure~2 presents radial dependences of other three absorption-line
indices resulting from our CCD observations with MPFS. We see
that the nucleus of NGC~2841 is also decoupled from the bulge
by increased strength
of the Fe lines; but the gradient of the Fe line strengths
in the bulge is quite negligible.
The {\it Fe5270} difference between the nucleus and the bulge
($0.9\pm0.3$\AA) corresponds to the metallicity difference
$\Delta [Fe/H] = 0.6\pm0.2$ dex
(by using the calibration of Worthey 1994). The slight depression
of the $H\beta$ absorption line in the central region of NGC~2841
results from
its contamination by emission. Overall, the small point-to-point
scatter in Fig.~2 confirms the high accuracy of the present index
determinations (r. m. s. is 0.1\,\AA, as mentioned above).

\section{Stellar and gaseous kinematics in the center of NGC~2841}

Figures~3 and 4 presents isovelocities for the gaseous (October 1994)
and stellar (February 1996) velocity fields in the center of NGC~2841
(the stellar
velocities are given with respect to the systemic velocity, because
the cross-correlation is made with the nuclear spectrum as a template).
For isovelocity
continuity, the derived velocity fields were smoothed by a gaussian
with $FWHM\approx 3\arcsec$; as for the stellar velocity
field, the smoothing has not affect the spatial resolution because of
rather poor seeing during the observations of February 1996, but the
maximum gas rotation velocity is in reality slightly higher than is
seen in Fig.~3.

\begin{figure}
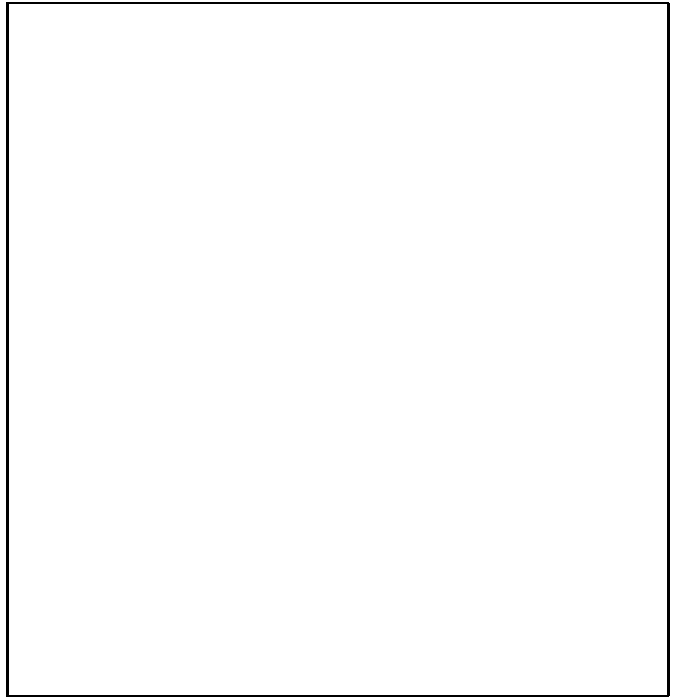

\picplace{9.2cm}
\caption[]{The isovelocities of ionized gas in the center of NGC~2841.
The cross marks the
photometric center of the galaxy. North is up, east is to the left.}
\end{figure}

\begin{figure}
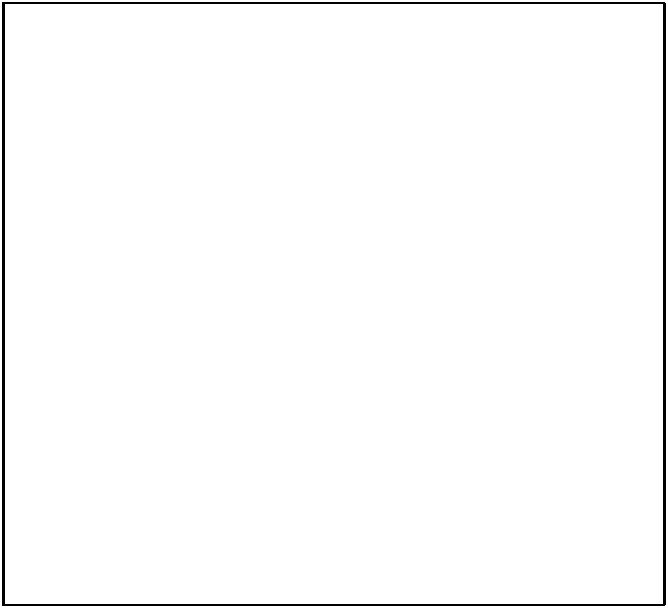

\picplace{8.0cm}
\caption[]{The isovelocities of stars in the center of NGC~2841 with
respect to the systemic velocity of the galaxy. The cross marks the
photometric center of the galaxy. North is up, east is to the left.}
\end{figure}

From the MPFS spectra, obtained in October 1994 at a rather good
spectral resolution (2 \AA) in the $H\alpha$ spectral region, we have
estimated a nuclear gas velocity dispersion as
170~$\mbox{km} \ \mbox{s}^{-1}$. The nuclear stellar velocity dispersion
is known to be of the same order: 220~$\mbox{km} \ \mbox{s}^{-1}$
(Delisle \& Hardy 1992). So we expected to find stellar and gaseous
rotation with comparable velocity amplitudes.
The rotations of both components are clearly seen in Figs. 3 and 4, but
their characters are quite different:
the ionized gas rotates perpendicular to
the stellar component. Another difference concerns the location of the
rotation velocity maxima: the gaseous component demonstrates two
prominent
velocity extremes -- one negative and one positive with respect to the
systemic velocity -- 3.8\arcsec\ from the center, and
the stellar component rotates like a solid body inside the radius
of 6\arcsec. If the dynamical center of the gaseous component lies
between the rotation-velocity extremes, it is shifted from the
photometric center by about 2\arcsec.
It is difficult to judge if the difference between the
dynamical and photometric center positions is real:
the extremes of the gas
velocity field are probably related to the edge of the nuclear
gaseous polar disk, because at 4.5\arcsec\ from the (dynamical) center
the zero-velocity line suddenly turns by $90\degr$, and the gas
further outward obviously rotates together with the stars.

Let us try to determine the orientations of the rotation axes more
exactly.
Under the assumption of circular rotation, an azimuthal dependence
of central line-of-sight velocity gradients must be a pure cosine
curve with a maximum at the orientation of the line of nodes $PA_0$: \\

\noindent
$dv_r/dr = \omega$ sin $i$ cos $(PA - PA_0)$, \\

\noindent
where $\omega$ is a deprojected angular rotation velocity and $i$ is an
inclination of the rotation plane. Two-dimensional velocity fields give
an unique opportunity to study such dependencies in detail. Figure~5
presents azimuthal dependencies of central line-of-sight velocity
gradients for the gaseous (unsmoothed) and stellar (smoothed) velocity
fields. The center is defined as photometric center, the range in radius
defined by two conditions -- to be inside the solid-body area and to
provide the maximum of measured directions -- is 2.8\arcsec--4\arcsec.

\begin{figure}
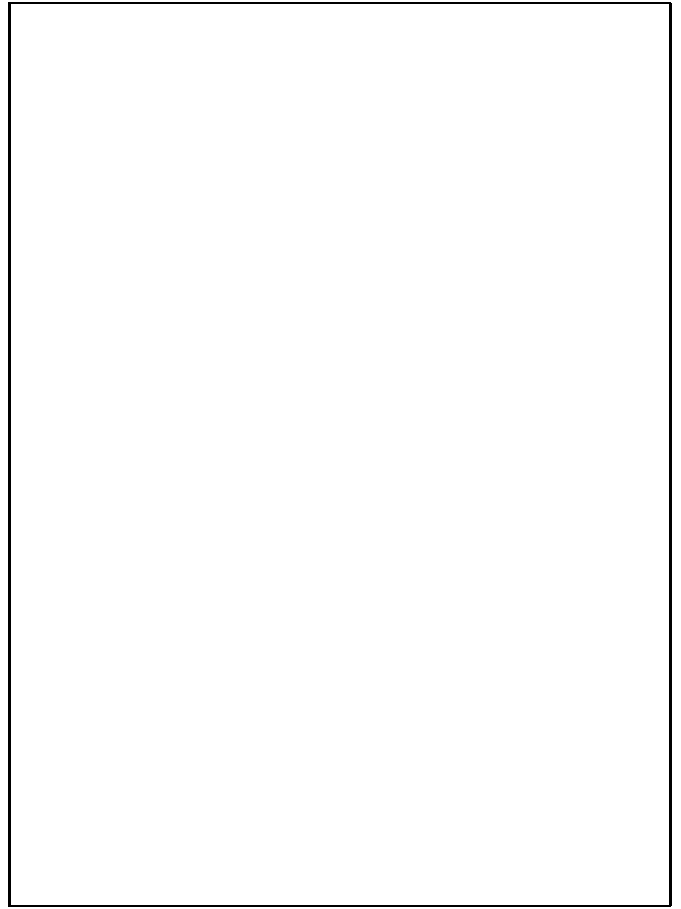

\picplace{12.0cm}
\caption[]{Azimuthal dependencies of the central line-of-sight velocity
gradients for NGC~2841 obtained with MPFS. The solid lines show cosine
curves fitted by the least-square method.}
\end{figure}

The cosine laws calculated by the least-square method are \\

\noindent
$dv_r/dr$ = [30.9 cos $(PA - 68\degr) - 0.8$]
$\mbox{km} \ \mbox{s}^{-1} \ \mbox{arcsec}^{-1}$, \\

\noindent
for the gaseous component and \\

\noindent
$dv_r/dr$ = [18.2 cos $(PA - 160\degr) - 0.2$]
$\mbox{km} \ \mbox{s}^{-1} \ \mbox{arcsec}^{-1}$, \\

\noindent
for the stellar component. We see that the rotation axes of both
components are indeed orthogonal. The difference of cosine curve
amplitudes may be related to different inclinations of rotation planes,
but more likely it results from the smoothing of the stellar velocity
field.
The overall shapes of the azimuthal dependencies in Fig.~5 resemble
cosine laws very closely, so we conclude that a circular rotation may be
the dominant component for both velocity fields, stellar and gaseous.
The low-amplitude wave oscillations of points around the cosine curve
in the $PA$ range $120\degr-280\degr$ in Fig.~5a may be real
because they are seen in both H$\alpha$ and
[NII]$\lambda 6583$; if this is the case, a small bar may be present in
the center of NGC~2841. This possibility will be discussed in the next
section.

\begin{figure}
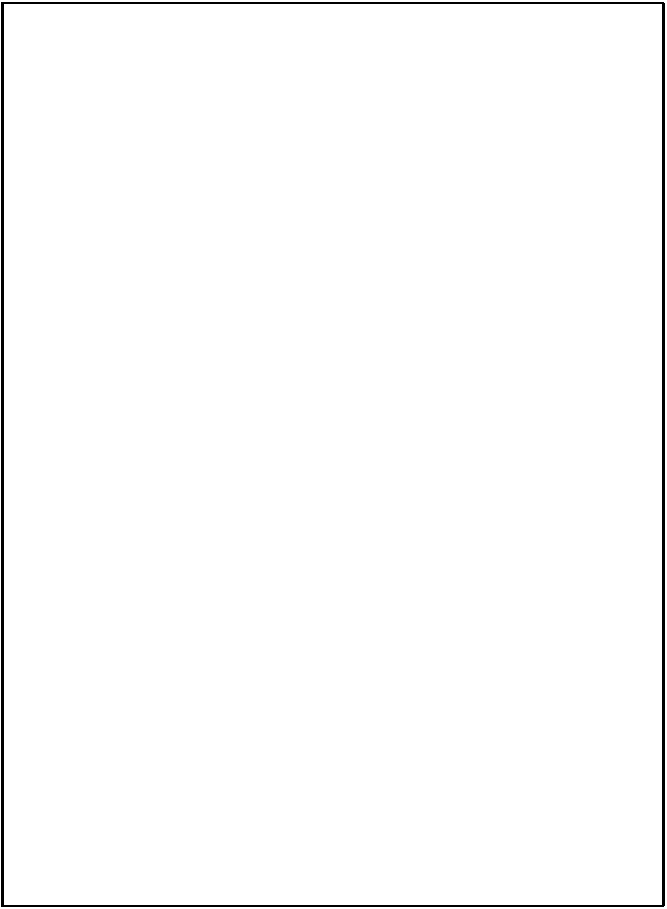

\picplace{12.0cm}
\caption[]{Azimuthal dependencies of the central line-of-sight velocity
gradients for NGC~2841 obtained with the long-slit spectrograph.
The solid lines show cosine curves fitted by the least-square method.}
\end{figure}

As we have mentioned above, four long-slit cross-sections were obtained
for NGC~2841 at the 1 m telescope in the position angles $152\degr$
(major axis), $242\degr$ (minor axis), $92\degr$, and $122\degr$
to verify the
properties of the stellar velocity field. A weak emission line
[OIII]$\lambda 5007$ has appeared to be quite measurable inside
$r\approx 5\arcsec$ in these spectra. Figure~6 presents an analog of
Fig.~5, but obtained with the long-slit spectrograph of the 1 m
telescope.
To compensate for a low count level, we have measured the spectra twice:
each individual row (a spatial element of 1.54\arcsec) and binning by
2 rows (a spatial element of 3.1\arcsec). The agreement of the two
measurements is satisfactory. The fitted cosine laws in Fig.~6 are \\

\noindent
$dv_r/dr$ = [24.2 cos $(PA - 61\degr) - 13$]
$\mbox{km} \ \mbox{s}^{-1} \ \mbox{arcsec}^{-1}$, \\

\noindent
for the gaseous component and \\

\noindent
$dv_r/dr$ = [23 cos $(PA - 152\degr) - 3.8$]
$\mbox{km} \ \mbox{s}^{-1} \ \mbox{arcsec}^{-1}$, \\

\noindent
for the stellar component. Though the number of directions involved
in Fig.~6 is much less than in Fig.~5, the orthogonality of the
rotation planes of the
gaseous and stellar components is fully confirmed. The angular rotation
velocities for gas and stars are almost equal here, the spatial
resolutions being equal, so our initial expectation
of the comparable rotation velocity amplitudes for the gas and stars
due to their comparable velocity dispersions is justified.

\begin{figure}
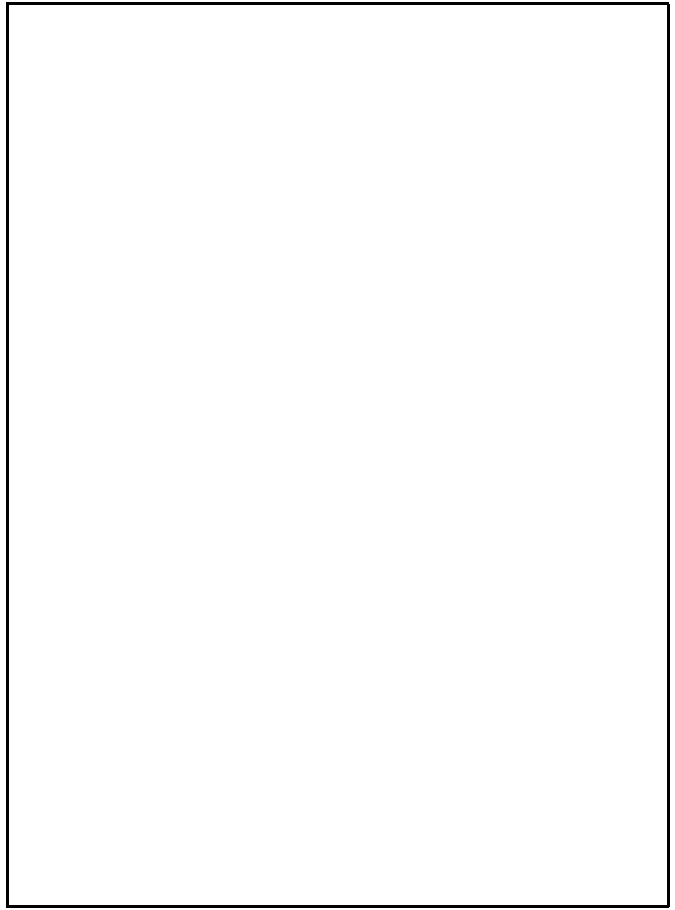

\picplace{12.0cm}
\caption[]{Comparison of our long-slit cross-sections along the major
($PA = 150\degr-152\degr$) and the minor ($PA = 60\degr-62\degr$) axes
with earlier measurements of Whitmore et al. (1984) and of Fillmore et
al. (1986).}
\end{figure}

Figure~7 presents a comparison of the major-axis and minor-axis
stellar velocity cross-sections with the earlier measurements
of Whitmore et al. (1984) and Fillmore et al. (1986).
To obtain a higher accuracy of velocity measurements,
we have cross-correlated separately two fragments of spectra --
in the wavelength ranges 4800--5320\,\AA\AA\ and 5320--5840\,\AA\AA, --
and each with
two spatial binnings, of 1.54\arcsec\ and of 3.1\arcsec, so we had four
measurements at each radius which being averaged have allowed us
to estimate
intrinsic errors in the means shown in Fig.~7. An independent check
of the wavelength scale accuracy was made by measuring the night-sky
emission line [OI]$\lambda5577$. After leaving aside systematic
velocity differences of 20 $\mbox{km} \ \mbox{s}^{-1} $ (major axis)
and 10 $\mbox{km} \ \mbox{s}^{-1} $ (minor axis), the agreement
of our measurements with previous ones may be accepted as satisfactory,
though one must keep in mind that their slits were much narrower than
ours.

\begin{figure}
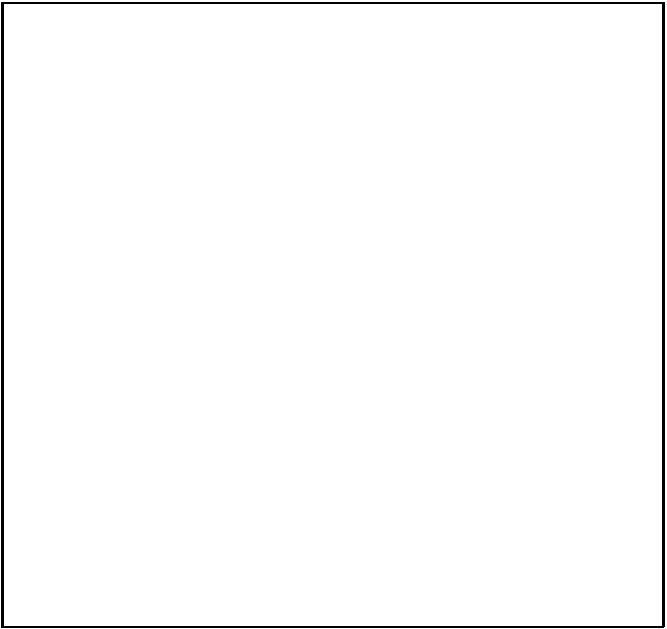

\picplace{8.3cm}
\caption[]{Long-slit line-of-sight stellar velocity profiles obtained
at the 1 m telescope. Fat straight lines indicate fits to the
observational data in the radius ranges of 0\arcsec--3.5\arcsec
and (only for $PA$ 92\degr and 242\degr) of 4.5\arcsec--10\arcsec.}
\end{figure}

The results for the all four position angles are presented in Fig.~8.
The curves have been shifted in
velocity to reach a maximum symmetry for the distributions
to the left and to the right from the nucleus. To show how successful
this procedure is, in Fig.~8 we present, in addition to the real
profiles (dark symbols), the curves mirrored about $r=0$
(light symbols). In three cross-sections out of four,
the nucleus has a line-of-sight velocity
systematically higher than the calculated systemic velocity
(the excess is from 20 to 40 $\mbox{km} \ \mbox{s}^{-1}$).
It may be explained by an uncertainty of slit positioning with respect
to the kinematical center of the galaxy due to a finite (large enough)
slit width or by the disagreement between the kinematical and
photometic centers mentioned above. But another effect seems
to be more interesting: in two cross-sections, $PA=92\degr$ and
$PA=242\degr$, there are indications of a counterrotating component
in the radius range of 5\arcsec--12\arcsec. Some low-amplitude motions
were earlier measured along the minor axis without any comments
(Whitmore et al. 1984; Fillmore et al. 1986), but after
symmetrization it appears that we deal with systematic
solid-body rotation (projected rotation velocity at $r_{pr} = 6\arcsec$
along the minor axis of the galaxy is 40 $\mbox{km} \ \mbox{s}^{-1}$)
of which the rotation axis does not coincide with that of more inner
and more outer parts of the galaxy. In Fig.~8, $P.A.=92\degr$, this
counterrotating (conditionally counterrotating because the
projected velocity along the minor axis is not zero) component is seen
even better. It lies between $r=4.5\arcsec$ and $r=10\arcsec$ and
demonstrates a slow solid-body rotation -- slower than that of the
kinematically decoupled core inside $r\approx 3\arcsec$. It could be
a projection effect; alternatively, the component could really be
dynamically hotter than the nucleus.

\begin{figure}
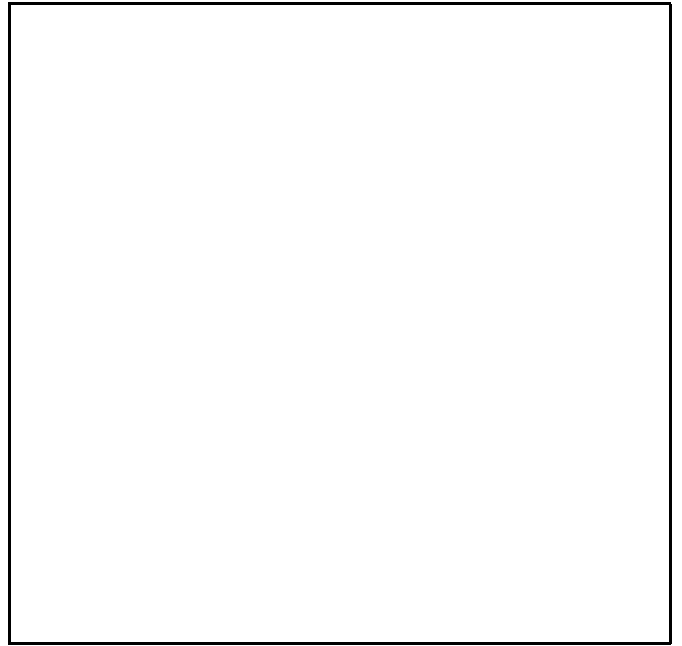

\picplace{8.5cm}
\caption[]{Composite rotation curve of the center of NGC~2841 calculated
by using the data obtained with MPFS and with the long slit. The solid
curve show fitting by a third-order polynomial. Arrows mark the
locations of "old" (mentioned by Whitmore et al.1984,
and by Fillmore et al. 1986) local
rotation velocity maximum and of the "new" one found in this paper.}
\end{figure}

Finally, Fig.~9 presents a rotation curve of the stellar component
of NGC~2841 which is calculated over the MPFS and the long-slit data.
The two-dimensional field shown in Fig.~4 was fitted by a circular
rotation model; the best parameters defining the orientation
of the plane of rotation
-- $PA_0 = 155\degr$ and $i=60\degr$ -- were used to calculate
the rotation curve shown in Fig.~9. Besides this curve,
two long-slit
cross-sections closest to the major axis were reduced
to the plane of the galaxy with the same orientation parameters.
All three curves show rather good agreement. The local
rotation-velocity maximum at $r=4.5\arcsec$ reported earlier by
Whitmore et al. (1984) and Fillmore et al. (1986) is seen but looks
far less prominent than in the data of our precursors. Instead, a higher
velocity maximum has appeared at $r=6\arcsec-10\arcsec$.
We will not to stress the problem
of the exact location of the local maximum of the rotation velocity.
It is more important that the analysis of the stellar velocity field
reveals that the central region of NGC~2841 is kinematically distinct,
and the radius of this kinematically decoupled core --
$r=3\arcsec-4\arcsec$ -- is only an upper limit because of poor
seeing (of order of 3\arcsec--4\arcsec\ too).

\section{May be a nuclear bar?}

\begin{figure}
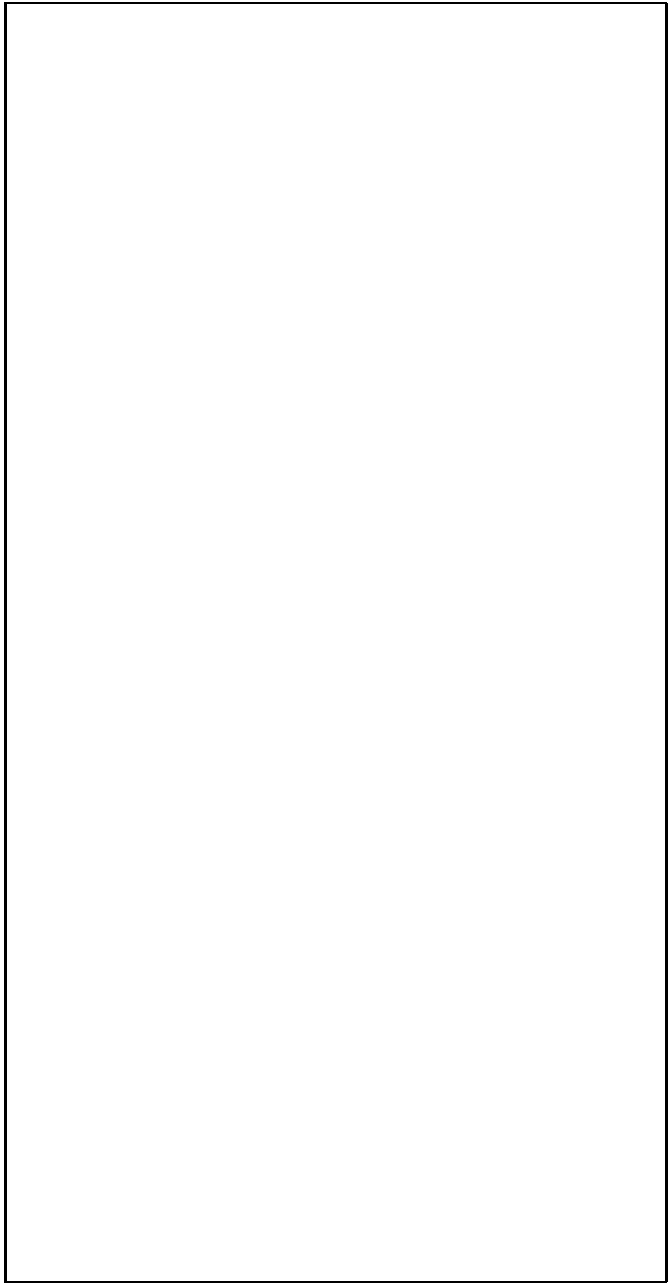

\picplace{17.0cm}
\caption[]{Radial dependencies of axial ratio, major axis position
angle, and boxiness derived from isophotal analysis of the central
part of NGC~2841.}
\end{figure}

Both visible effects, the counterrotation and the local velocity
maxima, can in
reality be artifacts produced by a triaxial compact structure
in the center of NGC~2841. The best way to verify if there is
a nuclear bar in the galaxy is to trace the orientation of isophotes
towards the center. If the structure is axisymmetrical, the position
angle of isophote major axis would be constant over all the radii;
if there is a bar, the isophote major axis would turn (the only
exception -- when a bar is aligned with the line of nodes -- can
be excluded because such a configuration results in a plateau on the
line-of-sight velocity profile along the major axis which
is not observed in NGC~2841).

Fig.~10 presents results of fitting the isophotes
obtained with MPFS by ellipses. Also in the same manner we have fitted
isophotes of an NGC~2841 image registered by the (unrepaired) WFPC
of the Hubble Space Telescope on September 23/24, 1992
(the exposure time was 260+260 sec, the filter was F555W,
the program No. 3912 of S. M. Faber);
the image was taken from the HST archive.
The characteristics of the outermost isophotes were obtained by
using the image of the galaxy from the Digital Sky Survey provided
by the SKYVIEW service. All three MPFS exposures have
given identical estimates of isophote ellipticities.
Though the HST image shows systematically larger ellipticities (by
about 0.1) than the MPFS images, the overall trend demonstrated by all
the data is monotonous, without any signs of increased ellipticity
in the center which must take place in the case of bar presence.
The $PA$ dependencies on radius look less definitive.
All three data sets obtained with MPFS
are quite consistent at $r=5\arcsec-6\arcsec$: here we have measured
$<PA$(major axis)$>=141.5\degr\pm1\degr$. But the orientation of the
WFPC isophotes in the same radius range is $PA=152\degr$, and Varela
et al. (1996) give $147\degr$. So we cannot state with certainty
that the orientation of the innermost isophotes in NGC~2841
is exactly coincident
with the line of nodes; but the turn if it exists is rather small,
and in any case there is no nuclear bar perpendicular to the line
of nodes (so called "edge-on bar") which alone can mimic circumnuclear
rotation velocity maxima and produce a cosine curve $dv_r/dr$ vs. $PA$
with the maximum at the line of nodes (Fig.~5b and 6b).
The HST image, $34\arcsec \times 34\arcsec$ in the central
CCD frame, has also allowed us to estimate the boxiness of the central
isophotes up to the radius of about 15\arcsec. In the semi-major axis
range of 6\arcsec--12\arcsec\ the isophotes have a clear boxy shape
with the mean $(a_4/a)\times 100=-0.32\pm0.04$. Let us note that
this semi-major axis range (taking into account the ellipticity of 0.3)
corresponds exactly to the radius range where there is an indication
of a kinematically decoupled
bulge zone in Fig.~8 ($PA=92\degr$ and $242\degr$).

\begin{figure}
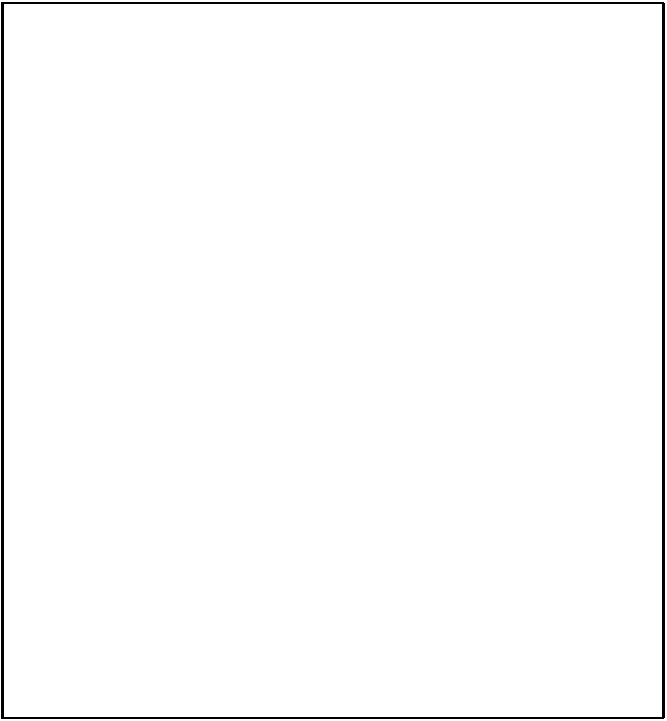

\picplace{9.5cm}
\caption[]{The image of the central part of NGC~2841 in the continuum
$\lambda6500$ (white isophotes) and in the emission line
[NII]$\lambda$6583 (grey-scaled contours) obtained with MPFS.}
\end{figure}

Figure~11 presents a composite image constructed for the central region
of NGC~2841 using the MPFS data:
the continuum at $\lambda 6500$ and the emission-line intensity for
[NII]$\lambda 6583$. While the continuum isophotes look like almost
perfect ellipses, without asymmetries or center shift and therefore
without any signs of dust presence, the nitrogen emission-line
distribution has a rather complex character. Keel (1983) reported that
emission in the center of NGC~2841 demonstrated a one-sided bar-like
distribution with a $PA$ of 105\degr. At a fainter level we
have obtained a distribution
resembling a two-sided bar, aligned with the line of nodes
and quite symmetrical with respect to the nucleus, but with some kind
of wings, or lobes, at a $PA\approx 30\degr$. Interestingly,
quite similar lobes with a similar orientation
($PA$ of about 50\degr) are seen on the outermost radio-continuum
isophotes ($r\approx 3\arcmin$) presented
by Condon (1987). As the lobes in Fig.~11 do not perfectly coincide
with the velocity extremes of the ionized gas (Fig.~3), we are not sure
that we see a polar gaseous disk overlapping the main gaseous disk
of the galaxy. But some structure perpendicular to the plane of the
galactic disk and seen in both ionized gas and radio continuum emission,
on scales from the circumnuclear area to the outermost parts of
NGC~2841, is obviously present.

\section{Discussion and conclusions}

In Sect.~3 we have seen that NGC~2841 has a chemically decoupled
nucleus. Its presence can be seen from the increased magnesium-line
and iron-line strengths.
Besides our observations, there are two more facts published earlier but
without special comments. Terndrup et al. (1994) present a $J-K$ color
profile for NGC~2841 which demonstrates a break of 0.4 mag between the
nucleus and the bulge at $r=3\arcsec$ and an absence of $J-K$ gradient
in the radius range 3\arcsec-100\arcsec. As the broad-band color $J-K$
is related mainly to the iron abundance, the similar behaviour of
the {\it Fe5270} and {\it Fe5335} profiles in our measurements is in
agreement with the results of Terndrup et al. (1994). Also,
Delisle \& Hardy (1992) have measured radial profiles of absorption
lines in the near-infrared spectral range; the CaII IR triplet profile
presented for NGC~2841 reveals a break of this absorption line index by
more than 1\AA\ between the nucleus and nearby bulge ($r < 8\arcsec$).
So we consider a chemical distinctness of the nucleus in NGC~2841 as
fully confirmed. Here we must only note that the estimates of abundance
breaks mentioned in the Sect.~3 -- 0.36 dex for the magnesium and
0.6 dex for the iron -- are lower limits because of uncertain age of
the nuclear stellar population. If the nuclear stellar population is
noticeably younger than the bulk of stars in the surrounding bulge,
the abundance difference estimates should be increased.

In Sect.~4 we have seen that NGC~2841 has a dynamically decoupled
nucleus and an unusual rotation of gas and stars in the outskirts of the
nucleus. The gas rotation is strictly perpendicular to that of
the nuclear stellar population, and also to the rotation of the global
neutral-hydrogen disk of the galaxy -- Rots (1980) gives an estimate of
the dynamical major axis of the large-scale HI distribution
$PA_0=155\degr$. There is an indication of a stellar
"counterrotating" component on the long-slit cross-sections
in directions far from the major axis, in the radius range
5\arcsec-12\arcsec. More probably, the mass distribution in the center
of NGC~2841 is near-axisymmetrical because of the almost constant
orientation of the continuum isophotes. In this case the angular
momentum of circumnuclear
gas and that of part of the stellar bulge are really decoupled, and
these components must have an external origin. We know one more Sb
galaxy with a similar set of properties -- NGC~7331. It has
a chemically decoupled stellar
nucleus (Sil'chenko et al. 1992), an axisymmetric nuclear mass
concentration traced by a decoupled fast solid-body rotation of
the ionized
gas (Afanasiev et al. 1989; Sil'chenko \& Vlasyuk 1992), and recently
a counterrotating stellar component was found in the bulge of NGC~7331
in the radius range 4\arcsec-15\arcsec\ (Prada et al. 1996).
The similarity is strong, and as NGC~7331 and 2841 are very nearby
galaxies, perhaps, a development of observational tools will lead
to a discovery of a whole class of such objects. What can be the
cause of such the set of properties?

As we have mentioned in the Introduction, a detection of chemically
decoupled nuclei in some galaxies with outer gas of obviously external
origin has given rise to an idea over the connection between galaxy
interaction and a phenomenon of decoupled nuclei. In the case of
NGC~2841 and 7331 the mechanism of galaxy interaction works quite well
too. There were numerous hints in the literature that interaction
perturbs large-scale gaseous disks of galaxies and provokes
intense gas inflows towards nuclei (see, for example, Tutukov \& Krugel
1995). A history of NGC~2841 may be the following. Some billions
years ago NGC~2841 had experienced an encounter with a smaller gas-rich
galaxy and had accreted some amount of its gas with decoupled momentum.
The gas of NGC~2841 itself was perturbed and in part settled into
the nucleus more quickly than the accreted external gas (due to more
effective dynamical friction in the large-scale disk of NGC~2841).
A star formation burst occured in the nucleus, and a secondary
chemically decoupled stellar population was produced. The accreted
external gas continued to settle towards the center through the bulge,
and in some vicinity of the nucleus
its density became high enough to begin another star formation burst;
the counterrotating part of the bulge had formed. Part of the accreted
gas has remained unlocked after this star formation burst and has
conserved its "strange" rotation up to date. Such a scenario may be
quite universal. In particular, in the frame of this hypothesis NGC~4826
where counterrotating gas lives only at $r> 1.5$ kpc and there is no
counterrotating stellar component yet, can be an early stage of
NGC~2841 and NGC~7331.

\begin{acknowledgements}
We are very grateful to the observers of the Special Astrophysical Observatory
RAS -- V.L. Afanasiev, S.N. Dodonov, and S.V. Drabek assisting us at the
6 m telescope. We are also grateful to the referee, Dr. R. Peletier,
whose comments have helped to make the paper more clear.
During the data analysis we have
used the Lyon-Meudon Extragalactic Database (LEDA) supplied by the
LEDA team at the CRAL-Observatoire de Lyon (France) and the NASA/IPAC
Extragalactic Database (NED) which is operated by the Jet Propulsion
Laboratory, California Institute of Technology, under contract with
the National Aeronautics and Space Administration. The work is partly based
on observations made with the NASA/ESA Hubble Space Telescope, obtained
from the data archive at the Space Telescope Science Institute, which is
operated by the Association of Universities for Research in Astronomy,
Inc., under NASA contract NAS 5-2655. 
We have used the software ADHOC developped at the Marseille Observatory.
The work was supported by the grants of the International Science Foundation
No. MMY300 and of the Russian Foundation for Basic Research No. 95-02-04480.
\end{acknowledgements}

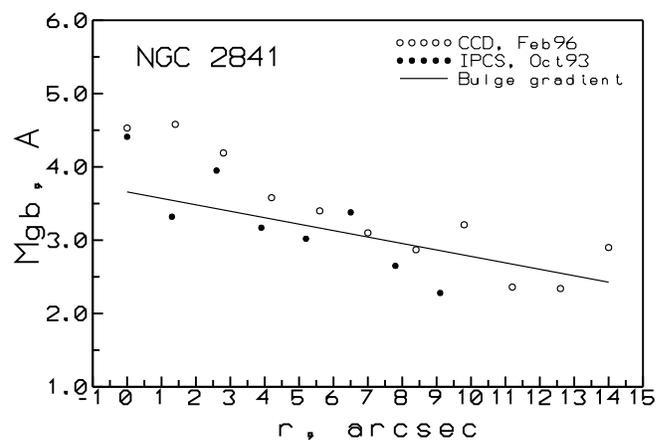
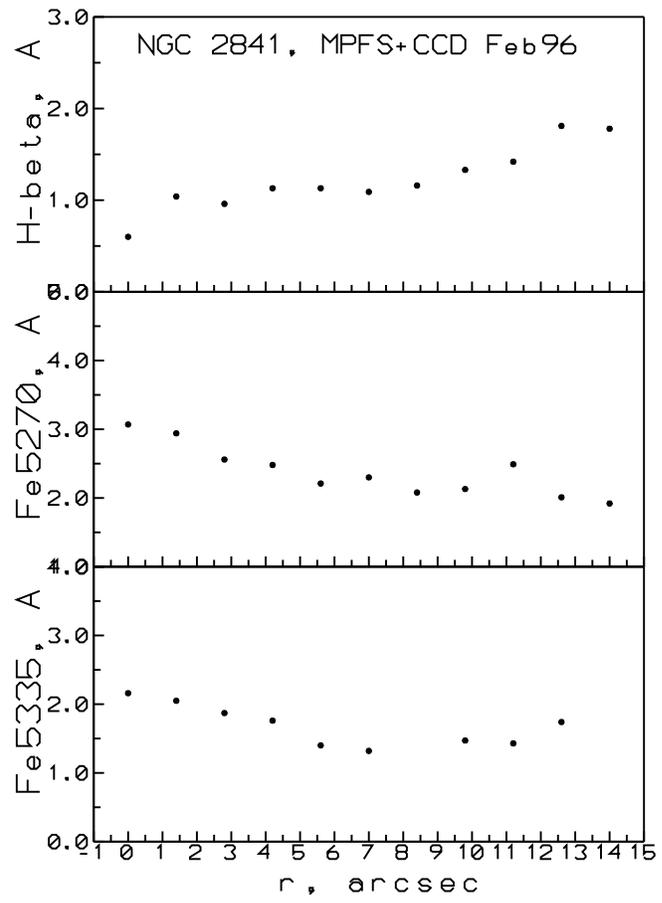
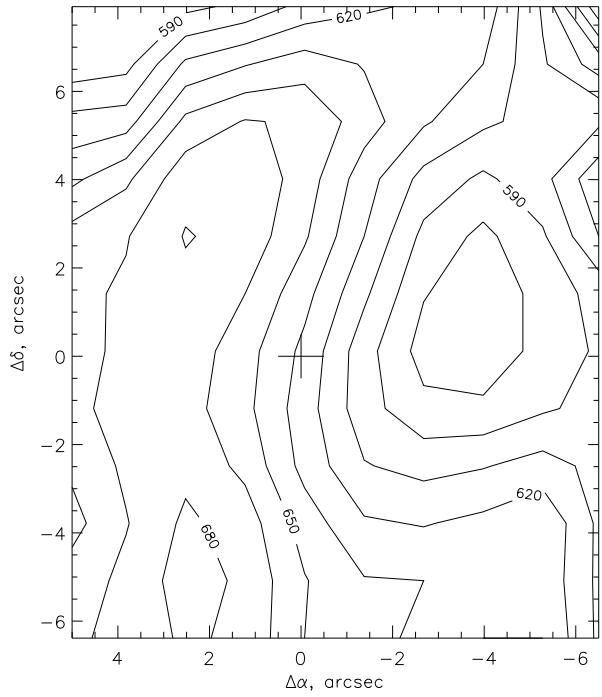
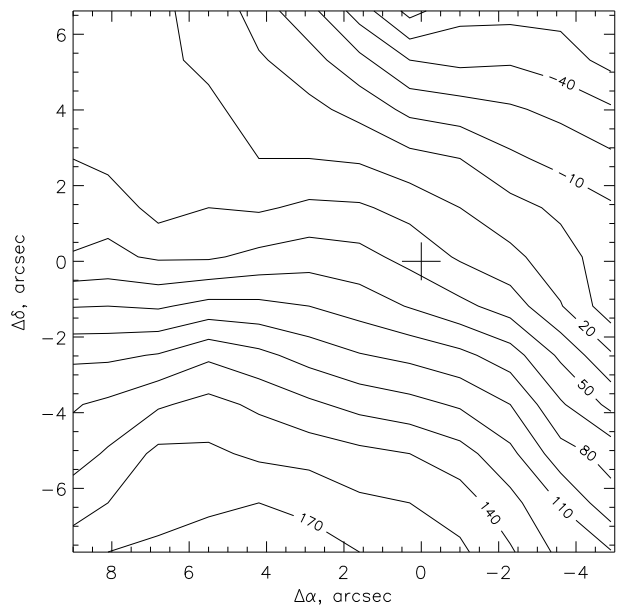
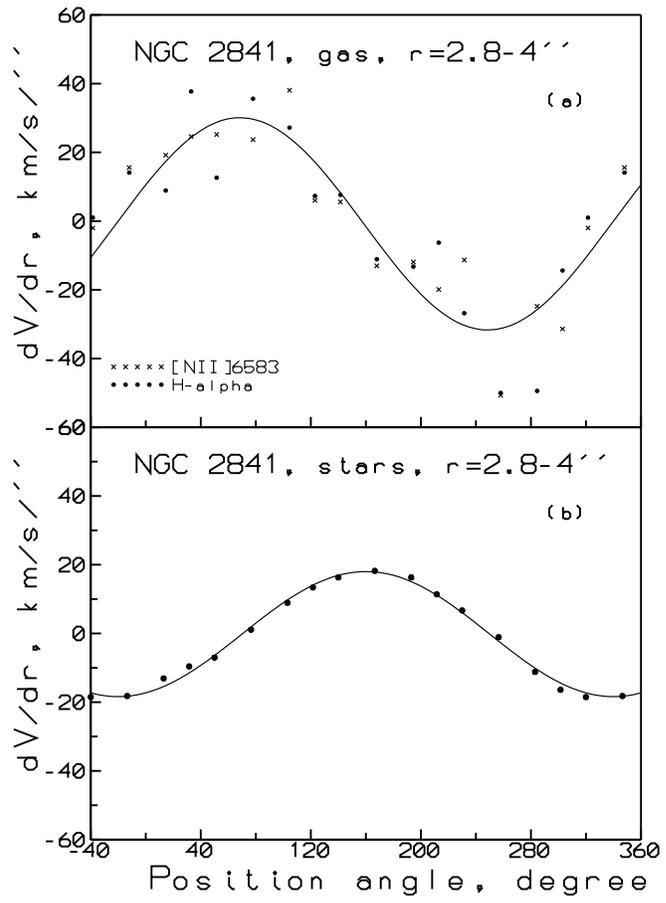
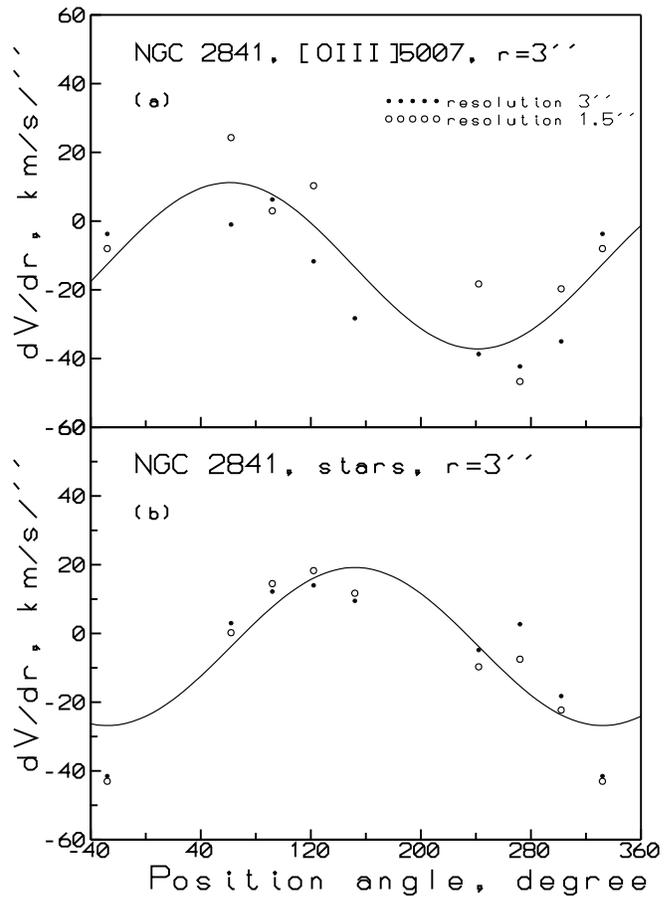
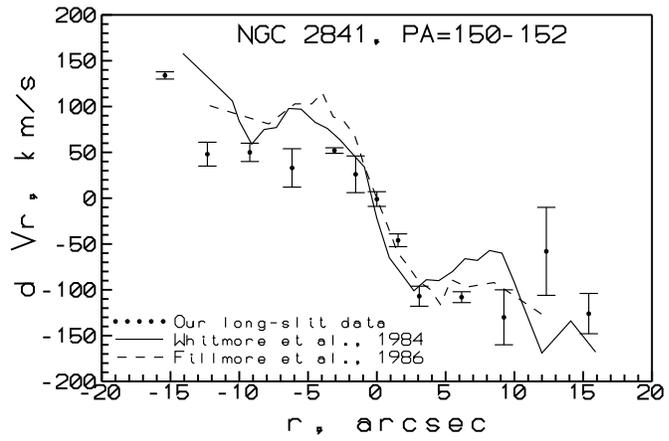
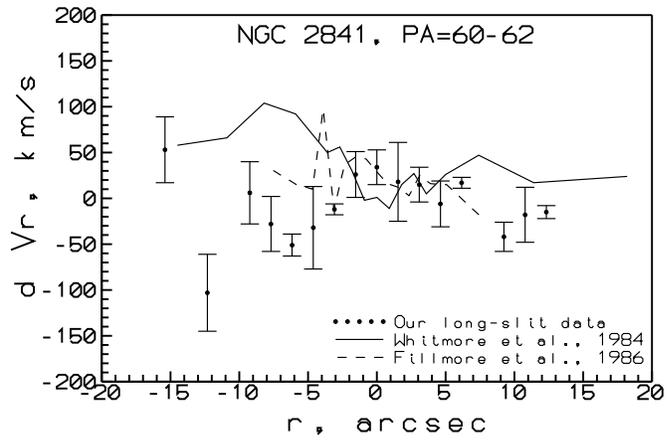
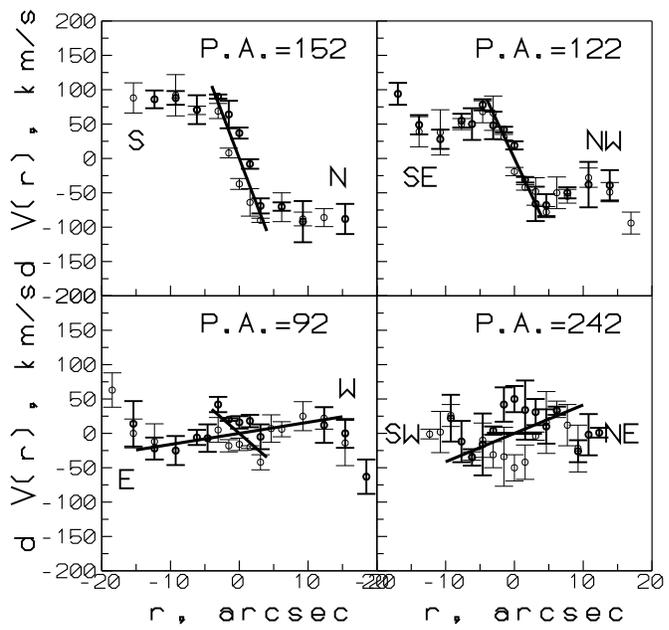
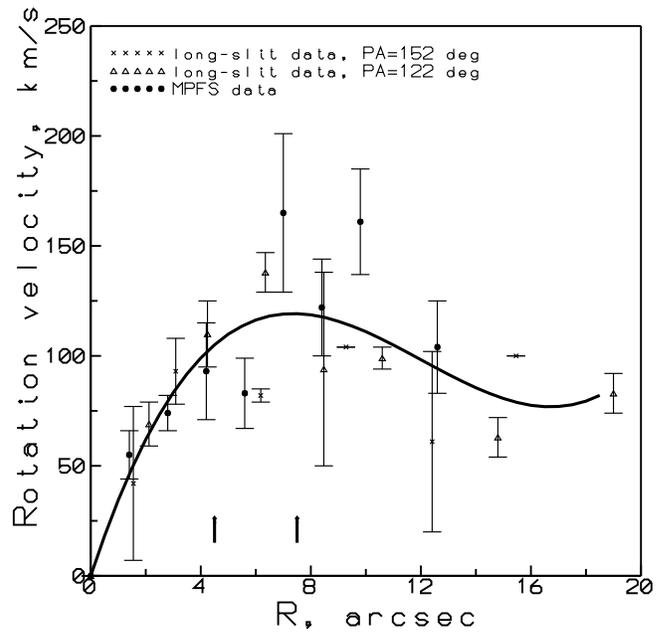
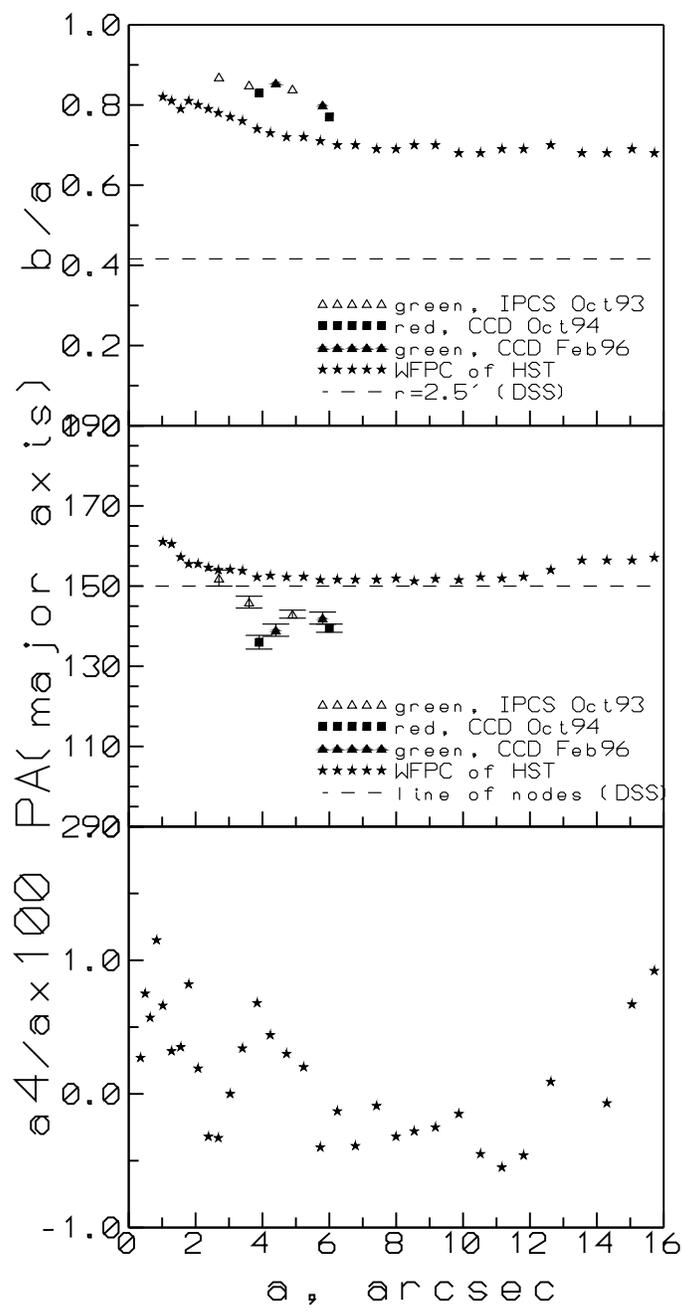
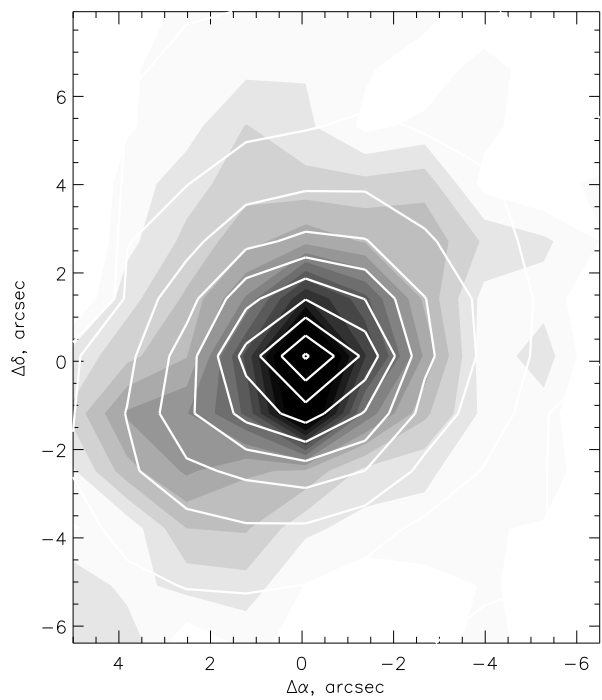
\end{document}